# Guidelines for reporting single-cell RNA-Seq experiments

Anja Füllgrabe[1,*], Nancy George[1], Matthew Green[1], Parisa Nejad[2], Bruce Aronow[3], Laura Clarke[1], Silvie Korena Fexova[1], Clay Fischer[4], Mallory Ann Freeberg[1], Laura Huerta[1,5], Norman Morrison[1], Richard H. Scheuermann[6], Deanne Taylor[7], Nicole Vasilevsky[8], Nils Gehlenborg[9], John Marioni[1], Sarah Teichmann[10], Alvis Brazma[1], Irene Papatheodorou[1,*]

[1] European Bioinformatics Institute, United Kingdom, Wellcome Genome Campus, Hinxton, Cambridgeshire, CB10 1SD, UK

[2] University of California, Santa Cruz, United States, 1156 High Street, Santa Cruz, CA, 95064, USA

[3] Cincinnati Children's Hospital Medical Center, United States, 240 Albert Sabin Way, Cincinnati, OH, 45229, USA

[4] University of California, Santa Cruz, United States

[5] Current address: Dotmatics Limited, United Kingdom, Bishop's Stortford CM23 2ND, UK

[6] J. Craig Venter Institute, United States; Department of Pathology, University of California San Diego, 4120 Capricorn Lane, La Jolla, CA, USA

[7] Children's Hospital of Philadelphia, United States

[8] Oregon Health & Science University, United States, 3181 SW Sam Jackson Park Road, Portland, OR 97239, USA

[9] Harvard Medical School, United States, Department of Biomedical Informatics, 10 Shattuck Street, Boston, MA 02115, USA

[10] Wellcome Sanger Institute, United Kingdom, Wellcome Genome Campus, Hinxton, Cambridgeshire, CB10 1SA, UK

[*] corresponding authors

# Abstract

Single-cell RNA-Sequencing (scRNA-Seq) has undergone major technological advances in recent years, enabling the conception of various organism-level cell atlassing projects. With increasing numbers of datasets being deposited in public archives, there is a need to address the challenges of enabling the reproducibility of such data sets. Here, we describe guidelines for a minimum set of metadata to sufficiently describe scRNA-Seq experiments, ensuring reproducibility of data analyses.

# Main

scRNA-Seq experiments have many advantages over the so-called bulk RNA-Sequencing and microarray experiments, as they allow researchers to study gene expression at an individual cell, rather than at tissue level[1]. As the scRNA-Seq technologies are maturing, these experiments are becoming increasingly high-throughput and widespread. Data from an estimated 1400 scRNA-Seq studies have been submitted to NCBI's Gene Expression Omnibus (GEO)[2], EMBL-EBI's ArrayExpress[3] and the European Nucleotide Archive[4] (ENA) over the recent years. Large collaborative efforts are taking off, such as the Human Cell Atlas (HCA)[5], which aims at uncovering the gene expression profiles of all human cell types, the Fly Cell Atlas (http://flycellatlas.org/), which has similar aims for Drosophila, or the Human Biomolecular Atlas Program (HuBMAP)[6], as well as organ-specific projects, such as the BRAIN Initiative Cell Census Consortium[7].

Meta-analyses combining data from independent scRNA-Seq studies have been performed, revealing that although overall conclusions from independent studies confirm each other, there are differences[8,9]. To ensure that the results of individual scRNA-Seq studies can be reproduced, to allow for reuse of data generated in such experiments, and more generally, to enable researchers to build on previous discoveries, it is important that the necessary minimal information about scRNA-Seq experiments are collected and preserved together with the respective data. The community agreement and publication of the Minimum Information About a Microarray Experiment (MIAME)[10] guidelines almost two decades ago, was a major milestone in the way functional genomics data has been reported and archived. For the last two decades, the functional genomics data archives at EBI and NCBI have been accepting microarray and bulk RNA-Seq datasets, but the emergence of protocols which can assay transcriptomics at single-cell resolution brings along new requirements.

There is a clear need to establish minimum standards for reporting data and metadata for the various scRNA-Seq assays. ArrayExpress and the HCA have published online guidelines for technical information required for scRNA-Seq data submissions, however these are not yet widely adopted community standards. Such standards would serve as the guiding principles and would ensure the reusability of the submitted datasets, guide the adaptation of existing archival resources and enable reproducibility of analysis by the wider scientific community. Here, we propose the minimum set of single-cell metadata terms, a checklist of information that is used to describe a single-cell assay in sufficient detail to enable analysis of the transcriptomic data. These guidelines are derived from the work on building and adapting community resources to archive and add value to such datasets, such as ArrayExpress, the HCA-Data Coordination Platform and the CIRM Stem Cell Hub.

Typical designs of single-cell transcriptomic experiments include the following steps: (a) single cell isolation; (b) addition of spike-in RNAs; (c) reverse transcription; (d) amplification; (e) library construction; (f) sequencing. Depending on the exact protocol followed, different types of metadata need to be recorded. Figure 1 shows an overview of the main steps that define the experimental workflows[11] for scRNA-Seq, with the variety of options used by different protocols.

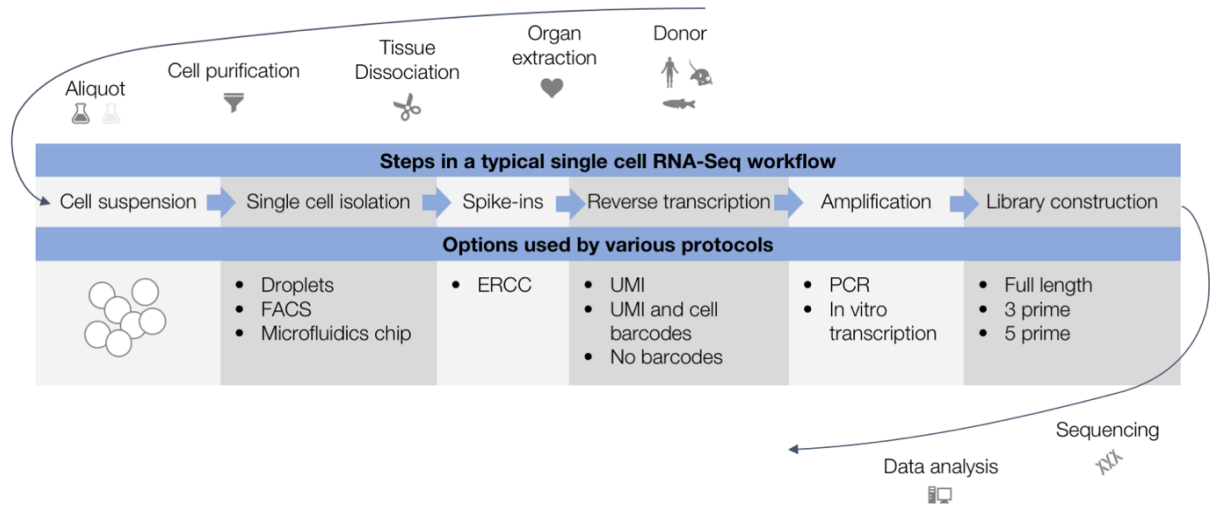

**Figure 1.** Typical design steps of single-cell transcriptomics experiments with examples for each step (ERCC: External RNA Controls Consortium; FACS: Fluorescence-Activated Cell Sorting; PCR: Polymerase Chain Reaction; UMI: Unique Molecular Identifier).

For example, the Smart-seq2 protocol[12], which is frequently used for single-cell transcriptomics, involves cell isolation using fluorescence-activated cell sorting (FACS) and the use of microwell plates to separate the single cells. Barcodes are typically not used during reverse transcription; amplification is done by PCR and libraries covering the full length of the sequences are constructed. Alternatively, microdroplet-based protocols such as the commonly used single-cell controller from 10x Genomics use droplets to encapsulate individual cells and barcode individual molecules during reverse transcription. It uses a 3- or 5-prime tag system during library preparation.

Taking the five main components of the MIAME and MINSEQE[10] experiment model, we can describe a single-cell sequencing experiment with a few additions and changes (Figure 2). Each component is equivalent to a main experimental step. We capture information describing each component and link it to the relevant protocols. Definitions of the individual components and a list of the single-cell specific attributes that are introduced can be found in the Supplementary Notes document. For each field we recommend using terms from a suitable ontology (like NCBI taxonomy[13] for species) or controlled vocabulary to prevent ambiguity. We refer to the Supplementary Notes document for examples of different implementations of the scRNA-Seq metadata guidelines.

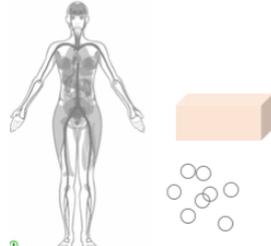

**Figure 2.** Overview of the five main components of a MINSEQE study and the new components specific for single-cell experiments (highlighted in yellow), with example attributes.

The adoption of MIAME guidelines by the scientific community, including the major scientific journals and public archives of functional genomics data almost two decades ago[14] was an important step towards enabling a widespread reuse of these data[15]. We strongly believe that now is the time to discuss and adopt similar guidelines for scRNA-Seq experiments, so that data generated in the growing number of these experiments are made available for reuse, meta-analysis and building of value-added gene expression resources. As single-cell transcriptomics are increasingly combined with imaging of tissue sections or quantification of surface proteins[16], future work will involve alignment of these standards with newly emerging techniques requiring new types of metadata. We also expect the standard to be expanded with single-cell genomic and epigenomic techniques (e.g. single-cell ATAC-seq), which are not covered here, to address different types of single cell assays more broadly. This would support the reuse and interoperability of various types of single-cell data and facilitate the development of atlases[17,18].

# Acknowledgements

This work was supported by the European Molecular Biology Laboratory, the Wellcome Trust Biomedical Resources grant Single Cell Gene Expression Atlas (108437/Z/15/Z), the California Institute for Regenerative Medicine (GC1R-06673-B), the Chan Zuckerberg Initiative DAF, an advised fund of the Silicon Valley Community Foundation (2018–182730), and by the NIH Common Fund, through the Office of Strategic Coordination/Office of the NIH Director (OT2OD026677).

# Author Contributions

A.F., N. George, M.G., P.N., B.A., S.K.F., C.F., M.A.F., L.H., D.T., N.V. and N.M. curated data and developed metadata standards; L.C., N.M., R.H.S., A.B., and I.P. supervised the project and engaged with the larger community; S.T. and J.M. provided support and evaluation of the standards and contributed to the manuscript; A.F., N. George, M.G., P.N., B.A., L.C.,

S.K.F., C.F., M.A.F., N. Gehlenborg, L.H., R.H.S, D.T., N.V., A.B. and I.P. wrote the manuscript.

## Competing Interests Statement

The authors declare no competing interests.

# Supplementary Note

# 1. Main components to describe a single-cell RNA-sequencing (scRNA-Seq) experiment

We reuse standards already defined by MIAME/MINSEQE[1] that include general information about the experiment and add new components and terms that are necessary for reproducibility and re-analysis of scRNA-Seq experiments.

Taking the five main components of the MINSEQE experiment model, we can describe a single-cell sequencing experiment with a few additions and changes (Supplementary Table 1). Each component is equivalent to a main experimental step. For each field we recommend using terms from a suitable ontology (like NCBI Taxonomy[2] for species, UBERON[3] for tissues, and Cell Ontology[4,5] for cell types) or controlled vocabulary. Protocols should ideally be hosted in a public repository such as protocols.io ([https://protocols.io](https://protocols.io)) and include a unique protocol identifier.

**Supplementary Table 1**. Descriptions of the components comprising the MINSEQE-derived guidelines for describing scRNA-Seq experiments.

| MINSEQE[+] Component | Description |
| --- | --- |
| **Biosource** | The characteristics of the biological sample material and experimental variables. Single cell experiments may have different starting points or several biosource transformation steps in the experimental flow, ranging from donor organism to final cell suspension. Any of these should be described with relevant attributes, as well as their interrelation. |
| **Single cell isolation** | Specific to single cell experiments is the cell isolation and singularisation information, for example the method how cells were dissected from tissue, sorted and enriched as well as visual quality checks prior to lysis. |
| **Library construction** | The library construction component comprises an overall description of the single-cell protocol, as well as more detailed information about the library preparation method, such as input fraction of biomolecule, primers used, strandedness, spike-in RNAs and information about barcodes for methods using multiplexing. |
| **Sequencing assay** | The sequencing assay refers to the sequencing of the library. It describes the sequencing mode and allows capture of technical replicate information such as whether the same library was sequenced multiple times or over different lanes and flow cells. |

| **Raw data** | The raw sequencing output, usually in FASTQ format. Additional attributes capture a file's identity (what type of read it contains), location and checksum. |
|---|---|
| **Processed data and post-analysis information** | Analysis results of a single-cell experiment, such as gene expression matrix with count values per gene and cell. Some processed data is useful for analysing single-cell experiments, such as a cell type definition that is determined as a result of a cell clustering or a list of cell barcodes. Although not within the "minimum" checklist, this information is helpful and recommended. |
| **Protocols** | Describes the procedures for transformation of the material or data from one experimental step to the next. |
| **Study and Publication** | General information about the study, such as a description and contributing authors, and a link to a published manuscript using the dataset. |

## 2. Single cell-specific attributes

In Supplementary Table 2, we describe the single cell-specific attributes that are captured for each component of the extended MINSEQE standard for scRNA-Seq datasets.

**Supplementary Table 2.** List of suggested single-cell specific attributes with definitions and significance. Significance levels are defined as 1: Required for analysis (mandatory); 2: Greatly improves data utility (essential); 3: Additional relevant information that can help improve analysis.

| Attribute | Description | Significance |
|---|---|---|
| **Biosource** | | |
| **Organism** | The species name of the donor organism. E.g. "Homo sapiens". | 1 |
| ***other sample attributes** | Any attributes describing the sample, cell suspension or biological materials they were derived from (e.g. donor age, sex, strain, genotype, disease/health status, tissue, cell type) as well as other experimental factors introduced during the experiment (e.g. compound treatment, sampling time point). | 2 |
| **Single cell isolation** | | |
| **Tissue dissociation** | The method by which tissues are dissociated into single cells in suspension. Examples are "proteolysis", "mesh passage", "fine needle trituration". | 3 |
| **Cell enrichment** | The method by which specific cell populations are sorted or enriched, e.g. "fluorescence-activated cell sorting (FACS)". | 3 |
| **Enrichment markers** | Description of the specificity markers used to isolate cell populations, e.g. "CD45+". | 2 |
| **Single cell isolation** | The method by which individual cells, nuclei, or another portion are separated for individual barcoding. Examples are "FACS", "microfluidics", "manual selection", "droplet-based cell isolation". | 2 |
| **Single cell entity** | The type of single cell entity derived from isolation protocol e.g. "whole cell", "nucleus", "cell-cell multimer", "spatially encoded cell barcoding". | 2 |
| **Single cell quality metric** | For plate/well-based methods information about the well contents may be available. This can either be measured by visual inspection prior to cell lysis or defined by known parameters such as wells with several cells or no cells. This can be captured at a high level, e.g. "OK" or "not OK", or with more specificity such as "debris", "no cell", "doublet", "50 cells". | 2 |
| **Cell number** | For droplet experiments, the number of cells used as input for the library preparation, e.g. "5000". | 2 |
| **Single cell identifier** | A unique identifier given to each individual cell, e.g. "cell 1". | 2 |
| **Library construction** | | |
| **Library construction** | The library construction method (including version) that was used, e.g. "Smart-Seq2", "Drop-Seq", "10X v3". | 1 |

| Input molecule | This is the specific fraction of biological macromolecule from which the sequencing library is derived, e.g. "polyA RNA". | 2 |
|---|---|---|
| Primer | The type of primer used for reverse transcription, e.g. "oligo-dT" or "random" primer. This allows users to identify content of the cDNA library input e.g. enriched for mRNA. | 2 |
| Amplification method | The method used to amplify RNA, e.g. "PCR" or "in vitro transcription". | 3 |
| Amplification cycles | The number of cycles used to multiply the RNA molecules. | 3 |
| End bias | The end of the cDNA molecule that is preferentially sequenced, e.g. 3/5 prime tag or end, or the full-length transcript. | 1 |
| Library strand | The strandedness of the library, whether reads come from both strands of the cDNA or only from the first (antisense) or the second (sense) strand. | 1 |
| Spike in | Spike-in RNAs are synthetic RNA molecules with known sequence that are added to the cell lysis mix. e.g. the External RNA Controls Consortium (ERCC) spike in set is commonly used in single-cell experiments. | 2 |
| Spike in dilution or concentration | The final concentration or dilution (for commercial sets) of the spike in mix, e.g. "1:40,000" | 2 |
| **Sequencing assay** | | |
| Library layout | If the library was sequenced in single-end or paired-end mode. | 2 |
| Instrument model | The manufacturer and model of the sequencing machine, e.g. "Illumina HiSeq 2500". | 2 |
| Technical replicate group or library reference | A common term for all runs/files belonging to the same cell or library. We suggest to use a stable sample accession from a biosample archive like BioSamples database[6]. Alternatively, the library ID can be referenced from which the files were generated. | 1 |
| **Raw data files and sequences** | | |
| UMI barcode read | The type of read that contains the UMI barcode: index1/index2/read1/read2. | 1 |
| UMI barcode offset | The offset in sequence of the UMI identifying barcode. E.g. "16". | 1 |
| UMI barcode size | The size of the UMI identifying barcode. E.g. "10". | 1 |
| Cell barcode read | The type of read that contains the cell barcode: index1/index2/read1/read2. | 1 |
| Cell barcode offset | The offset in sequence of the cell identifying barcode. E.g. "0". | 1 |
| Cell barcode size | The size of the cell identifying barcode. E.g. "16". | 1 |
| cDNA read | The type of read that contains the cDNA read: index1/index2/read1/read2. | 1 |
| cDNA read offset | The offset in sequence for the cDNA read. E.g. "0". | 1 |
| cDNA read size | The size of the cDNA read. E.g. "98". | 1 |
| Sample barcode read | The type of read that contains the sample barcode: index1/index2/read1/read2. | 3 |
| **Sample barcode offset** | The offset in sequence of the sample identifying barcode. E.g. "0". | 3 |
| **Sample barcode size** | The size of the sample identifying barcode (bp). E.g. "8". | 3 |
| **Read 1 file** | The name of the file that contains read 1. E.g. "File1_R1.fastq.gz". | 1 |
| **Read 2 file** | The name of the file that contains read 2 (for paired-end sequencing). E.g. "File1_R2.fastq.gz". | 1 |
| **Index 1 file** | The name of the file that contains index read 1. E.g. "File1_I1.fastq.gz". | 1 |
| **Index 2 file** | The name of the file that contains index read 2. E.g. "File1_I2.fastq.gz". | 1 |
| **Checksum** | Result of a hash function calculated on the file content to assert file integrity. Commonly used algorithms are MD5 and SHA-1. E.g. "ab21a9c0f2007890fe2fbe48df0519f9". | 2 |
| **White list barcode file** | A file containing the known cell barcodes in the dataset. | 3 |
| **Cell- and sample-associated information derived from data analysis** | | |
| **Inferred cell type** | Post analysis cell type declaration based on expression profile or known gene function identified by the performer, e.g. if a pool of cells are sequenced with the purpose of identifying new cell types or sub-populations. Per cell, in addition to sample of origin metadata, any number of lineage, class, and subclass attributes that serve to group cells within the sample or set of samples can be declared. E.g. "type II bipolar neuron". | 2 |
| **Post analysis cell/well quality** | Performer defined measure of whether the read output from the cell was included in the sequencing analysis. For example, cells might be excluded if a threshold percentage of reads did not map to the genome or if pre-sequencing quality measures were not passed. E.g. "pass" or "fail". | 2 |
| **Protocols** | | |
| **Single cell isolation protocol** | How were single cells separated into a single-cell suspension? | 3 |
| **Nucleic acid library construction protocol** | How was the cDNA library constructed? If a kit was used name, manufacturer, catalogue number, lot and date of manufacture should be included. | 3 |
| **Nucleic acid sequencing protocol** | How was the sequencing done, were any special parameters used? How many sequencing runs were done? Expected read lengths. | 3 |
| **Data analysis protocol** | What workflow was followed to derive results from raw, primary, and secondary data? How were cell types inferred? | 3 |

# 3. Implementations

## Tabular Format

MAGE-TAB is a spreadsheet-based format, developed to represent experimental metadata compliant with MIAME/MINSEQE[7]. A simple tabular format, such as MAGE-TAB is well suited for capturing information at the sample (i.e. cell) level. In a sequential way, it follows the sample through the stages of the experimental process, linking different entities from tissue/cell through different protocols for RNA extraction, sequencing library preparation and sequencing to the resulting data files. It is also able to express complex relationships between samples and data, such as different types of technical replicates.

The Gene Expression team at EMBL-EBI has modified the original MAGE-TAB format, both the Investigation Description Format (IDF) and the Sample and Data Relationship Format (SDRF), to capture single-cell specific information. The SDRF has been expanded with single-cell specific terms as "Comment" columns. This allows users to provide all the information required for processing at the single-cell level within a single metafile. Moreover, scRNA-Seq datasets are annotated with the experiment type ontology term "RNA-seq of coding RNA from single cells" (EFO_0005684) in the IDF. In contrast, bulk RNA-Seq datasets are annotated as "RNA-seq of coding RNA" (EFO_0003738), allowing specific retrieval of only scRNA-Seq datasets, only bulk RNA-Seq datasets or both. In order to describe a scRNA-Seq experiment, the IDF should specify the following protocols: 1) sample collection protocol (EFO_0005518), 2) single cell isolation protocol (EFO term pending), 3) single cell nucleic acid library construction protocol (EFO term pending) and 4) single cell sequencing protocol (EFO_0008439). A complete list of MAGE-TAB fields used for archival in ArrayExpress and analysis in Single Cell Expression Atlas can be found at https://github.com/ebi-gene-expression-group/sc-metadata-fields.

*An example describing a Smart-Seq2 experiment in Drosophila melanogaster*

We have applied our minimum metadata standards to the Smart-Seq2 dataset generated by Li et al. in 2017[8]. The IDF corresponding to study GSE100058, including a curated description of each protocol, can be downloaded here ftp://ftp.ebi.ac.uk/pub/databases/microarray/data/atlas/sc_experiments/E-GEOD-100058/E-GEOD-100058.idf.txt.

The SDRF has been adapted to include single-cell specific information in order to enable re-use and interpretation of single-cell experiments. Supplementary Table 3 summarises the metadata representation of a sample from the GSE100058 dataset. Here, in the case of the plate-based Smart-Seq2 library preparation protocol, one sample corresponds to one individual cell. The single-cell specific information entities are added as columns to the SDRF table to allow individual annotation of each cell.

**Supplementary Table 3.** Overview of the metadata fields applied to a sample of Smart-Seq2 experiment GSE100058.

| SDRF field | Ontology ID | Example value | Ontology ID |
|---|---|---|---|
| **biomaterial** | | | |
| Source Name | | GSM2670685 | |
| Characteristics[organism] | OBI_0100026 | Drosophila melanogaster | NCBITaxon_7227 |
| Characteristics[organism part] | EFO_0000635 | brain | UBERON_0000955 |
| Characteristics[sex] | PATO_0000047 | mixed sex population | EFO_0001271 |
| Characteristics[developmental stage] | EFO_0000399 | pupal stage | FBdv_00005349 |
| Characteristics[genotype] | EFO_0000513 | GH146-GAL4 | |
| Characteristics[cell type] | EFO_0000324 | olfactory projection neuron | |
| Characteristics[single cell quality] | | OK | |
| Characteristics[well information] | | 1 cell | |
| **single cell isolation** | | | |
| Comment[single cell isolation] | | FACS | EFO_0009108 |
| **single cell library preparation** | | | |
| Comment[input molecule] | | polyA RNA | OBI_0000869 |
| Comment[library construction] | OBI_0000711 | smart-seq2 | EFO_0008931 |
| Comment[end bias] | | none | |
| Comment[primer] | | oligo-dT | |
| Comment[spike in] | | none | |
| Comment[library_strand] | | not applicable | |
| Comment[library_layout] | | PAIRED | |
| Comment[library_source] | | TRANSCRIPTOMIC SINGLE CELL | |
| Comment[library_strategy] | | OTHER | |
| Comment[library_selection] | | other | |
| Comment[ENA_experiment] | | SRX2921940 | |
| **sequencing** | | | |
| Comment[instrument_model] | | Illumina NextSeq 500 | |
| **sequencing data file** | | | |
| Comment[ENA_run] | | SRR5687154 | |

*An example describing a 10x experiment in Mus musculus*

In contrast to plate-based technologies, where each and every cell can be characterised, the high throughput of droplet-based single-cell technologies, such as Drop-Seq or 10x, makes it impractical to annotate cells individually in the SDRF. Moreover, the cells in the same sequencing reaction have identical metadata for the attributes that are known before the data analysis. Therefore, to describe a droplet-based experiment, information is captured in the SDRF at the level of the sequencing library, with each library typically containing a few thousand cells.

E-MTAB-6429 is a 10x experiment comparing HPV16 E7 oncogene expressing mouse keratinocytes versus wild type cells[9]. The SDRF has four samples, describing the four sequencing libraries (2 wild type and 2 transgenic keratinocyte samples). Supplementary Table 4 lists the metadata fields used to annotate a sample in E-MTAB-6429. Here, additional fields are required to describe how the data (i.e. sequencing reads) for each cell and molecule can be retrieved. The IDF file can be downloaded from https://www.ebi.ac.uk/arrayexpress/files/E-MTAB-6429/E-MTAB-6429.idf.txt.

**Supplementary Table 4**. Overview of the metadata fields applied to a sample of 10x experiment E-MTAB-6429.

| SDRF field | Ontology ID | Example value | Ontology ID |
|---|---|---|---|
| **biomaterial** | | | |
| Source Name | | SAMEA5043064 | |
| Characteristics[organism] | OBI_0100026 | Mus musculus | NCBITaxon_10090 |
| Characteristics[strain] | EFO_0005135 | C57BL/6 | EFO_0004472 |
| Characteristics[sex] | PATO_0000047 | female | PATO_0000383 |
| Characteristics[age] | EFO_0000246 | 10 to 12 | |
| Characteristics[genotype] | EFO_0000513 | wild type genotype | EFO_0005168 |
| Characteristics[organism part] | EFO_0000635 | epidermis | UBERON_0001003 |
| Characteristics[sampling site] | EFO_0000688 | ear | UBERON_0001690 |
| Characteristics[cell type] | EFO_0000324 | keratinocyte | CL_0000312 |
| **single cell isolation** | | | |
| Comment[single cell isolation] | | 10x | EFO_0008995 |
| **single cell library preparation** | | | |
| Comment[input molecule] | | polyA RNA | OBI_0000869 |
| Comment[library construction] | OBI_0000711 | 10xV1 | |
| Comment[end bias] | | 3 prime tag | |
| Comment[primer] | | oligo-dT | |
| Comment[spike in] | | none | |

| Comment[cDNA read] | read1 | |
|---|---|---|
| Comment[cDNA offset] | 0 | |
| Comment[cDNA size] | 98 | |
| Comment[UMI barcode read] | read2 | |
| Comment[UMI barcode offset] | 0 | |
| Comment[UMI barcode size] | 1 | |
| Comment[cell barcode read] | index1 | |
| Comment[cell barcode offset] | 0 | |
| Comment[cell barcode size] | 14 | |
| Comment[sample barcode read] | index2 | |
| Comment[sample barcode offset] | 0 | |
| Comment[sample barcode size] | 8 | |
| Comment[library_strand] | not applicable | |
| Comment[library_layout] | PAIRED | |
| Comment[library_source] | TRANSCRIPTOMIC SINGLE CELL | |
| Comment[library_strategy] | OTHER | |
| Comment[library_selection] | PolyA | |
| Comment[ENA_experiment] | ERX2862695 | |
| **sequencing** | | |
| Comment[instrument_model] | Illumina NextSeq 500 | |
| **sequencing data file** | | |
| Comment[ENA_run] | ERR2856110 | |

## HCA JSON format

The Human Cell Atlas (HCA) metadata standard is represented and validated by a JSON schema. The schema stipulates each attribute's description, data type, programmatic name, human-readable name, example values, and guidelines for usage. By putting this information within the schema itself, everything needed to easily interpret and project metadata is available in one source. For example, the schema can be used to build tables and forms for data contributor input, to validate submitted metadata against the HCA metadata standard, to map to other metadata standards, and to convert metadata from one format to another (e.g. tabulated to nested JSON).

Similar to MAGE-TAB, the entities described by JSON schema are designed to reflect the experimental design. Biomaterial entities of various types (e.g. donor, cell suspension) are arranged to reflect the experimental workflow. Process and protocol information is captured to describe the transformation of one biomaterial into another, for example transforming a specimen into a cell suspension through dissociation. Entity linking is relied upon to further add expressive context to the metadata, for example linking two different specimens to the same donor to reflect biological replicate samples.

Biomaterials are also linked to data and supplementary data files which have their own set of associated metadata fields. Schema modularity affords fungibility to accurately represent experimental workflow, while the JSON schemas themselves ensure that the fields are tightly controlled and validated to maintain a high metadata standard.

Examples are given in Supplementary Table 5 (Smart-Seq2 technology) and Supplementary Table 6 (10x technology).

**Supplementary Table 5**. Metadata captured for a Smart-Seq2 HCA dataset. The subset of fields supplied by the data contributors are shown; unused fields are not shown. Contributor metadata data is not shown as there are many contributors per project. Link: https://data.humancellatlas.org/explore/projects/ae71be1d-ddd8-4feb-9bed-24c3ddb6e1ad.

| Metadata Field | Metadata Value | Scope |
|---|---|---|
| **Project** | | |
| Project short name | Healthy and type 2 diabetes pancreas | Per project |
| Project title | Single-cell RNA-Seq analysis of human pancreas from healthy individuals and type 2 diabetes patients | |
| Project description | We used single-cell RNA-Sequencing to generate transcriptional profiles of endocrine and exocrine cell types of the human pancreas. Pancreatic tissue and islets were obtained from six healthy and four T2D cadaveric donors. Islets were cultured and dissociated into single-cell suspension. Viable individual cells were distributed via fluorescence-activated cell sorted (FACS) into 384-well plates containing lysis buffer. Single-cell cDNA libraries were generated using the Smart-Seq2 protocol. Gene expression was quantified as reads per kilobase transcript and per million mapped reads | |
| | (RPKM) using rpkmforgenes. Bioinformatics analysis was used to classify cells into cell types without knowledge of cell types or prior purification of cell populations. We revealed subpopulations in endocrine and exocrine cell types, identified genes with interesting correlations to body mass index (BMI) in specific cell types and found transcriptional alterations in T2D.  Complementary whole-islet RNA-Seq data have also been deposited at ArrayExpress under accession number E-MTAB-5060 (http://www.ebi.ac.uk/arrayexpress/experiments/E-MTAB-5060). | |
| Supplementary links | https://www.ebi.ac.uk/gxa/sc/experiments/E-MTAB-5061/Results | |
| INSDC project accession | ERP017126 | |
| ArrayExpress accession | E-MTAB-5061 | |
| INSDC study accession | PRJEB15401 | |
| Authors | Segerstolpe A\|\|Palasantza A\|\|Eliasson P\|\|Andersson EM\|\|Andreasson AC\|\|Sun X\|\|Picelli S\|\|Sabirsh A\|\|Clausen M\|\|Bjursell MK\|\|Smith DM\|\|Kasper M\|\|Ammala C\|\|Sandberg R | Per publication; zero or more per project |
| Publication title | Single-Cell Transcriptome Profiling of Human Pancreatic Islets in Health and Type 2 Diabetes | |
| DOI | 10.1016/j.cmet.2016.08.020 | |
| PMID | 27667667 | |
| Publication URL | https://europepmc.org/abstract/MED/27667667 | |
| Name | - | Per contributor; one or more per project |
| Email | - | |
| Phone number | - | |
| Institute | - | |
| Laboratory | - | |
| Address | - | |
| Country | - | |
| Project role | - | |
| ORCID ID | - | |
| Corresponding contributor | - | |
| Grant ID | 648842 | Per funder; one or more per project |
| Organization | European Research Council | |
| **Donor** | | |
| Donor ID | H1 | Per donor |
| Donor description | Normal Donor 1 | |
| NCBI taxon ID | 9606 | |
| Genus species | Homo sapiens | |
| Genus species ontology | NCBITaxon:9606 | |
| Is living | no | |
| Biological sex | male | |
| Known disease(s) | normal | |
| Known disease(s) ontology | PATO:0000461 | |
| Organism age | 43 | |
| Organism age unit | year | |
| Organism age unit ontology | UO:0000036 | |
| Development stage | human adult stage | |
| Development stage ontology | HsapDv:0000087 | |
| Body mass index | 30.8 | |
| Test results | HbA1c 5.0% | |
| **Specimen** | | |
| Specimen ID | H1_pancreas | Per specimen, one or more per donor |
| Specimen description | Pancreas from normal donor H1 | |
| Organ | pancreas | |
| Organ ontology | UBERON:0001264 | |
| Organ part | islet of Langerhans | |
| Organ part ontology | UBERON:0000006 | |
| Purchased specimen Manufacturer | Prodo Laboratories Inc (Irvine, CA, USA) | |
| **Cell suspension** | | |
| Cell suspension ID | AZ_A7 | Per cell suspension; one or more per specimen |
| Cell suspension description | AZ_A7 cell | |
| INSDC ID | ERS1348479 | |
| BioSample ID | SAMEA4437030 | |
| Plate label | AZ | |
| Well label | A7 | |
| Cell quality | OK | |
| Estimated cell count | 1 | |
| **Sequence file** | | |
| File name | AZ_A7.fastq.gz | Per file; one or two per cell suspension (single or paired end) |
| File format | fastq.gz | |
| Read index | read1 | |
| Read length | 43 | |
| INSDC run | ERR1630022 | |
| INSDC experiment | ERX1700355 | |
| **Dissociation protocol** | | |
| Dissociation protocol ID | dissociation_protocol_1 | Zero or more per project |
| Dissociation protocol description | Islets were dissociated into single-cell suspension and viable individual cells were distributed by FACS into 384-well plates containing lysis buffer. | |
| Publication DOI | 10.1016/j.cmet.2016.08.020 | |
| Dissociation method | fluorescence-activated cell sorting | |
| Dissociation method ontology | EFO:0009108 | |
| **Library preparation protocol** | | |
| Library preparation protocol ID | library_preparation_protocol_1 | Zero or more per project |
| Library preparation protocol description | Single-cell RNA-Seq libraries were generated as described in Picelli et al., 2014, Full-length RNA-Seq from single cells using Smart-Seq2, Nature Protocols. | |
| Publication DOI | 10.1038/nprot.2014.006 | |
| Input nucleic acid molecule | polyA RNA | |
| Input nucleic acid molecule ontology | OBI:0000869 | |
| Nucleic acid source | single cell | |
| Spike-in dilution | 40000 | |
| Spike-in kit Retail name | External RNA Controls Consortium (ERCC) | |
| Spike-in kit Manufacturer | Ambion, Life Technologies | |
| Library construction approach | Smart-seq2 | |
| Library construction approach ontology | EFO:0008931 | |
| Library construction kit Retail name | Nextera XT kit | |
| Library construction kit Manufacturer | Illumina | |
| End bias | full length | |
| Primer | poly-dT | |
| Strand | unstranded | |
| **Sequencing protocol** | | |
| Sequencing protocol ID | sequencing_protocol_1 | Zero or more per project |
| Sequencing protocol description | Libraries were sequenced on an Illumina HiSeq 2000, generating 43 bp single-end reads. | |
| Publication DOI | 10.1038/nprot.2014.006 | |
| Instrument manufacturer model | Illumina HiSeq 2000 | |
| Instrument manufacturer model ontology | EFO:0004203 | |

| Paired end | no |
| Sequencing approach | full length single cell RNA sequencing |
| Sequencing approach ontology | EFO:0008441 |

**Supplementary Table 6**. Metadata captured for a 10x HCA dataset. The subset of fields supplied by the data contributors are shown; unused fields are not shown. Contributor metadata data is not shown as there are many contributors per project. Link: https://data.humancellatlas.org/explore/projects/005d611a-14d5-4fbf-846e-571a1f874f70.

| Metadata Field | Metadata Value | Scope |
|---|---|---|
| **Project** | | |
| Project short name | HPSI human cerebral organoids | Per project |
| Project title | Assessing the relevance of organoids to model inter-individual variation | |
| Project description | The purpose of this project is to assess the relevance of pluripotent stem cell-derived cerebral and liver organoids to recapitulate the variation in cell-type specific gene expression programs between individuals. Towards this aim, we will generate reference atlases of the developing cortex and liver from multiple individuals, derive iPSC lines from these same individuals, and determine if inter-individual gene expression variation is recapitulated in cerebral and liver organoids from the same individual from which we have reference maps. In parallel we will assess the genetic contribution to variability between organoids from different iPSCs of multiple human individuals that are available in existing iPSC resources (e.g. HipSci). | |
| Contact name | - | Per contributor; one or more per project |
| Email | - | |
| Phone number | - | |
| Institution | - | |
| Laboratory | - | |
| Address | - | |
| Country | - | |
| Project role | - | |
| ORCID ID | - | |
| Corresponding contributor | - | |
| Grant ID | Not provided | Per funder; one or more per project |
| Organization | Chan Zuckerberg Initiative | |
| **Donor** | | |
| Donor ID | HPSI0214i-wibj | Per donor |
| Genus species | Homo sapiens | |
| Genus species ontology | NCBITaxon:9606 | |
| NCBI taxon ID | 9606 | |
| BioSample ID | SAMEA2398911 | |
| Biological sex | female | |
| Organism age | 55-59 | |
| Organism age unit | year | |
| Organism age unit ontology | UO:0000036 | |
| Development stage | adult | |
| Development stage ontology | HsapDv:0000087 | |
| Ethnicity | European, White, British | |
| Ethnicity ontology | HANCESTRO:0462 | |
| Known disease(s) | normal | |
| Known disease(s) ontology | PATO:0000461 | |
| Is living | yes | |
| **Specimen** | | |
| Specimen ID | HPSI0214i-wibj_skin | Per specimen; one or more per donor |
| Genus species | Homo sapiens | |
| Genus species ontology | NCBITaxon:9606 | |
| NCBI taxon ID | 9606 | |
| BioSample ID | SAMEA2397844 | |
| Organ | skin of body | |
| Organ ontology | UBERON:0002097 | |
| Organ part | skin epidermis | |
| Organ part ontology | UBERON:0001003 | |
| Known disease(s) | normal | |
| Known disease(s) ontology | PATO:0000461 | |
| **Cell line** | | |
| Cell line ID | HPSI0214i-wibj_2 | Per cell line; one or more per specimen |
| Genus species | Homo sapiens | |
| Genus species ontology | NCBITaxon:9606 | |
| NCBI taxon ID | 9606 | |
| BioSample ID | SAMEA2627567 | |
| Known disease | normal | |
| Known disease ontology | PATO:0000461 | |
| Cell type | pluripotent stem cell | |
| Cell type ontology | CL:0002248 | |
| Model organ | Stem cell | |
| Model organ ontology | CL:0000034 | |
| Catalog number | 77650057 | |
| Catalog URL | [http://www.hipsci.org/lines/#/lines/HPSI0214i-wibj_2](http://www.hipsci.org/lines/#/lines/HPSI0214i-wibj_2) | |
| Cell line type | induced pluripotent | |
| Cell viability method | Growth to confluence post-thaw | |
| Passage number | 32 | |
| Growth medium | mTeSR1 | |
| Feeder layer type | feeder-free | |
| Drug treatment | Cells were cultured in presence of Penicillin and Streptomycin | |
| Mycoplasma testing method | PCR | |
| Mycoplasma testing results | pass | |
| Date established | 2014-10-24T00:00:00Z | |
| **Organoid** | | |
| Organoid ID | Org_HPSI0214i-wibj_2_1 | Per organoid; one or more per cell line |
| Genus species | Homo sapiens | |
| Genus species ontology | NCBITaxon:9606 | |
| NCBI taxon ID | 9606 | |
| BioSample ID | SAMEA5643382 | |
| Organ | brain | |
| Organ ontology | UBERON:0000955 | |
| Organoid age | 62 | |
| Organoid age unit | day | |
| Organoid age unit ontology | UO:0000033 | |
| Organoid embedded in matrigel | yes | |
| Organoid growth environment | suspension | |
| **Cell suspension** | | |
| Cell suspension ID | HPSI_organoids_pooled_1 | Per cell suspension; one or more per organoid |
| Genus species | Homo sapiens | |
| Genus species ontology | NCBITaxon:9606 | |
| NCBI taxon ID | 9606 | |
| BioSample ID | SAMEA5643396 | |
| INSDC Sample accession | ERS3447460 | |
| Selected cell type | neural cell | |
| Selected cell type ontology | CL:0002319 | |
| Total estimated cell count | 6316 | |
| **Sequence file** | | |
| File name | GAC027_hOrg_HipSci_1_S5_L007_I1_001.fastq.gz | Per file; two, three, or four per cell suspension (index1 and index2 optional) |
| File format | fastq.gz | |
| Checksum | 79b4def79acf215ca440a11570f1e97f | |
| Read index | index1 | |
| Lane index | 7 | |
| Read length | 8 | |
| INSDC run accession | ERR3239100 | |
| INSDC experiment accession | ERX3266489 | |
| **iPSC induction protocol** | | |
| iPSC induction protocol ID | ipsc_induction_protocol_1 | Zero or more per project |
| iPSC induction protocol description | Fibroblasts are thawed, transduced using Cytotune 2.0 Sendai virus (containing the Yamanaka genes encoding transcription factors Oct4, Sox2, cMyc and Klf4) and maintained until iPSC colony formation. Colonies are then picked and cultured to obtain a sizable yield of IPS cells, which are banked to a commercial grade standard. These banks then undergo quality checks to ensure the banks pass resuscitation tests and are free of mycoplasma. | |
| Document filename | hipsci-ipsc-pipeline.pdf | |
| Induction method | sendai virus | |
| Pluripotent vector removed? | yes | |
| iPSC induction kit Retail name | Cytotune 1.0 | |
| iPSC induction kit Manufacturer | Thermofisher | |
| Pluripotency test | HipSci Pluri test | |
| **Differentiation protocol** | | |
| Differentiation protocol ID | Org_Lanc_2014 | Zero or more per project |
| Differentiation protocol description | Generation of cerebral organoids from human pluripotent stem cells. | |
| Publication DOI | 10.1038/nprot.2014.158 | |
| Differentiation method | embryoid bodies | |
| Target pathway | RHO, ROCK | |
| Differentiation validation method | immunostaining | |
| Retail name | ROCK inhibitor Y27632 | |
| Small molecules | Vitamin A (retinoic acid) | |
| **Dissociation protocol** | | |
| Dissociation protocol ID | Cerebral_organoid_dissociation | Zero or more per project |
| Document filename | Dissociation_protocol_130-092-628.pdf | |
| Dissociation method | Papain-based enzymatic dissociation | |
| Dissociation method ontology | EFO:0009128 | |
| Retail name | Neural Tissue Dissociation Kit | |
| Catalog number | 130-092-628 | |
| Manufacturer | Miltenyi Biotec | |
| **Library preparation protocol** | | |
| Library preparation protocol ID | 10x_3'_library_preparation | Zero or more per project |
| Library preparation protocol description | 10x Chromium single cell 3' v2 library preparation | |
| Document filename | CG00052_SingleCell3_ReagentKitv2UserGuide_RevE.pdf | |
| Nucleic acid source | single cell | |
| Input nucleic acid molecule | polyA RNA | |
| Input nucleic acid molecule ontology | OBI:0000869 | |
| Library construction approach | 10X 3' v2 sequencing | |
| Library construction approach ontology | EFO:0009899 | |
| End bias | 3 prime tag | |
| Primer | poly-dT | |
| Strand | first | |
| Cell barcode-containing read | Read 1 | |
| Cell barcode offset | 0 | |
| Cell barcode length | 16 | |
| UMI barcode-containing read | Read 1 | |
| UMI barcode offset | 15 | |
| UMI barcode length | 10 | |
| Library construction kit Retail name | 10X Chromium Single Cell 3' Solution v2 Chemistry | |
| Library construction kit Manufacturer | 10X Genomics | |
| **Sequencing protocol** | | |
| Sequencing protocol ID | 10x_scRNASeq | Zero or more per project |
| Sequencing protocol description | 10x RNA sequencing | |

| Document filename | CG00052_SingleCell3_ReagentKitv2UserGuide_RevE.pdf |
|---|---|
| Instrument manufacturer and model | Illumina HiSeq 2500 |
| Instrument manufacturer and model ontology | EFO:0008565 |
| Paired end | yes |
| Sequencing approach | tag based single cell RNA sequencing |
| Sequencing approach ontology | EFO:0008440 |

## CIRM Tag Storm

The CIRM CESCG (California Institute for Regenerative Medicine Center of Excellence in Stem Cell Genomics) is an initiative comprised of two dozen labs working on a wide variety of samples and assays. Biomaterials undergo extensive transformation. One such example is a donor with a disease condition having samples taken, the samples dissociated and sorted, PBMCs (peripheral blood mononucleocytes) isolated and reprogrammed into induced pluripotent stem cells, those cells are then differentiated and made into model systems such as cerebral organoids. Assay snapshots are taken during various steps to see how the biomaterials respond to being transformed.

This complex biomaterial transformation is best captured utilizing a bottom-up approach to metadata, as labs have such differing experimental designs that it can be difficult to predict what information is necessary in a top-down schema. A subset of these metadata fields was initially developed as the Minimum Information about a Stem Cell Experiment (MISCE) metadata standard.  Subsequently, this metadata standard and schema has been modified as the CIRM Tag Storm format. Tag Storm is a human- and computer-readable format first developed by the UCSC Genome Browser.

```
/* tagStorm - stuff to parse and interpret a genome-hub metadata.txt
 * file, which is in a hierarchical format.  That is something like:
 *
 *      cellLine HELA
 *      lab ucscCore
 *
 *          target H3K4Me3
 *          antibody abCamAntiH3k4me3
 *
 *              file hg19/chipSeq/helaH3k4me3.narrowPeak.bigBed
 *              format narrowPeak
 *
 *              file hg19/chipSeq/helaH3K4me3.broadPeak.bigBed
 *              format broadPeak
 *
 *          target CTCF
 *          antibody abCamAntiCtcf
 *
 *              file hg19/chipSeq/helaCTCF.narrowPeak.bigBed
 *              format narrowPeak
 *
 *              file hg19/chipSeq/helaCTCF.broadPeak.bigBed
 *              format broadPeak
 *
 * The file is broken into stanzas, that are separated from each other
 * by blank lines. The blank lines can include whitespace, which is
 * ignored.  Multiple blank lines can separate stanzas with no
 * difference in meaning from a single blank line. Lines within a stanza
 * start with a word which is the field label.
 * A tab or a space (or multiple tabs or spaces) separate the label from
 * the field contents. The label can't contain spaces.
 *
 * The file is interpreted so that lower level stanzas inherit tags from
 * higher level ones. This file might be used as so:
 *    struct tagStorm *tags = tagStormFromFile("metadata.txt");
 *    struct hash *fileIndex = tagStormUniqueIndex(tags, "file");
```

```
 *    struct tagStanza *stanza = 
 * hashMustFindVal(fileIndex,"hg19/chipSeq/helaCTCF.broadPeak.bigBed");
 *    char *target = tagFindVal(stanza, "target");      // Target is CTCF
 *
 * Most commonly indentation is done one tab-character per indentation
 * level. Spaces may also be used. If spaces and tabs are mixed
 * sometimes you get surprises. The tab-stop is interpreted as happening
 * every 8 spaces.
 *
 */
```

# 4. Methods

Data submissions to ArrayExpress

ArrayExpress is one of the major databases for functional genomics experiments and receives various kinds of single cell-based datasets. These include different kinds of scRNA-Seq approaches, as well as single-cell DNA-Seq projects (e.g. single-cell ATAC-Seq, single-cell MNase-Seq).

The versatility of new technologies creates a demand to quickly adapt the metadata model and submission tool to new requirements. One challenge was to incorporate new file formats, such as the 10x Genomics file model that stores vital barcode sequences in a third or fourth FASTQ file, in addition to the usual two files for a paired-end sequencing run. These efforts lead to the creation of new tags in the SAM/BAM format specification to accommodate the single cell-specific barcodes together with the genomic/transcriptomic sequence.

For directly submitted ArrayExpress experiments, metadata is curated at the level of submission, which allows curators to collect information from the data provider. The interaction with the data submitters is valuable to not only fill missing bits of information in the existing metadata schema, but also to learn about new, specific components of a particular protocol that have not been considered before.

The Human Cell Atlas Data Coordination Platform of the Human Cell Atlas

The Human Cell Atlas (HCA) Data Coordination Platform (DCP) metadata standard and schema implementation is maintained following an iterative, community-based, data-driven approach. The HCA metadata standard is developed in a transparent and open manner so that the HCA community can participate in the process. Data generators producing data from new cellular-resolution technologies can request updates to the HCA metadata standard in order to better capture information about new data types, for example 10x sequencing-specific metadata fields. Data consumers can also request updates to the metadata standard to support analysis and visualization tool development. Community requests are reviewed by the HCA DCP Metadata Team who assess the value of the requested update to the entire HCA community and, in some cases, agree to adopt the change as part of the HCA metadata standard. The HCA DCP Metadata Team includes members from both the EMBL-EBI and UCSC, and it is have made the most significant contribution to the schema and commits (now totalling close to 3,800) are tracked in the HCA GitHub repository (https://github.com/HumanCellAtlas/metadata-schema).

Single cell technologies are a rapidly developing field. As such the HCA metadata standard needs to be able to adapt accordingly, with regular updates and a process for managing and tracking schema versions. By designing with this principle in mind, data requiring updates to the metadata standard can quickly be submitted to the HCA and made available to consumers.

We anticipate that the HCA will expand to support other data modalities including controlled access data, model organism data, data from disease cohorts, proteomic, metabolomics, genomic and possibly even data from genetically engineered biological samples. The HCA has therefore developed a flexible and modular data model that can be extended to new data types and modalities ([https://data.humancellatlas.org/metadata/design-principles/structure#motivation](https://data.humancellatlas.org/metadata/design-principles/structure#motivation)). By designing with the principle of flexibility in mind, the HCA metadata standard will be prepared to support all future data.

Discussion with the wider community

The reporting guidelines for scRNA-Seq experiments have been extensively discussed through a series of conference calls with the ArrayExpress, HCA, and Human BioMolecular Atlas Program (HuBMAP) communities. Original paper drafts and the metadata implementations of the different projects have been open for input and comparison during that time to support the derivation of the minimum, common terms required for reproducibility of scRNA-Seq experiments.